\documentclass[5p]{elsarticle}
\usepackage[english]{babel}
\usepackage[T1]{fontenc}
\usepackage{amsmath,amssymb}
\usepackage{graphicx}
\usepackage{hyperref}
\usepackage{natbib}
\usepackage{dcolumn}
\makeatletter
\providecommand{\doi}[1]{%
  \begingroup
    \let\bibinfo\@secondoftwo
    \urlstyle{rm}%
    \href{http://dx.doi.org/#1}{%
      doi:\discretionary{}{}{}%
      \nolinkurl{#1}%
    }%
  \endgroup
}
\makeatother

\biboptions{sort&compress}
\journal{Physics Letters B}
\begin{document}
    \begin{frontmatter}
    \author[1,2]{M. Holl}
    \author[1,2]{R. Kanungo}
    \author[3,4]{Z. H. Sun}
    \author[3,4]{G. Hagen}
    \author[5,6]{J.A. Lay}
    \author[5,6]{A. M. Moro}
    \author[2]{P. Navr\'atil}
    \author[3,4]{T. Papenbrock}
    \author[2]{M. Alcorta}
    \author[2]{D. Connolly}
    \author[2]{B. Davids}
    \author[7]{A. Diaz Varela}
    \author[2]{M. Gennari}
    \author[2]{G. Hackman}
    \author[2]{J. Henderson}
    \author[8]{S. Ishimoto}
    \author[7]{A. I. Kilic}
    \author[2]{R. Kr\"{u}cken}
    \author[2,9]{A. Lennarz}
    \author[9]{J. Liang}
    \author[10]{J. Measures}
    \author[11,12]{W. Mittig}
    \author[2]{O. Paetkau}
    \author[9]{A. Psaltis}
    \author[13]{S. Quaglioni}
    \author[1]{J. S. Randhawa}
    \author[2]{J. Smallcombe}
    \author[13]{I. J. Thompson}
    \author[2,14]{M. Vorabbi}
    \author[2,15]{M. Williams}

    \address[1]{Astronomy and Physics Department, Saint Mary's University, Halifax, Nova Scotia, B3H 3C3, Canada}
    \address[2]{TRIUMF, Vancouver, BC, V6T 2A3, Canada}
    \address[3]{Physics Division, Oak Ridge National Laboratory, Oak Ridge, TN 37831, USA}
    \address[4]{Department of Physics and Astronomy, University of Tennessee, Knoxville, TN 37996, USA}
    \address[5]{Departamento de F\'isica At\'omica, Molecular y Nuclear, Universidad de Sevilla, Apartado 1065, E-41080 Sevilla, Spain}
    \address[6]{Instituto Interuniversitario Carlos I de F\'isica Te\'orica y Computacional (iC1), Apdo.~1065, E-41080 Sevilla, Spain}
    \address[7]{Department of Physics, University of Guelph, Guelph, ON, N1G 2W1, Canada}
    \address[8]{High Energy Accelerator Research Organization (KEK), Ibaraki 305-0801, Japan}
    \address[9]{McMaster University, Hamilton, ON, L8S 481, Canada}
    \address[10]{Department of Physics, University of Surrey, Guildford, Surrey, GU2 7XH, United Kingdom}
    \address[11]{National Superconducting Cyclotron Laboratory, Michigan State University, East Lansing, MI 48824-1321, USA}
    \address[12]{Department of Physics and Astronomy, Michigan State University, East Lansing, MI 48824-1321, USA}
    \address[13]{Lawrence Livermore National Laboratory, P.O. Box 808, L-414, Livermore, California 94551, USA}
    \address[14]{National Nuclear Data Center, Bldg. 817, Brookhaven National Laboratory, Upton, New York 11973-5000, USA}
    \address[15]{Department of Physics, University of York, York YO10 5DD, United Kingdom}

\title{Proton inelastic scattering reveals deformation in $^8$He}
    
\begin{abstract}
  A measurement of proton inelastic scattering of $^8$He at $8.25A$~MeV at TRIUMF shows a resonance
  at 3.54(6)~MeV with a width of 0.89(11)~MeV. The energy of the state is in good agreement with 
  coupled cluster and no-core shell model with continuum calculations, with the latter successfully
  describing the measured resonance width as well. Its differential cross section analyzed with phenomenological collective excitation form factor and microscopic coupled reaction channels framework consistently reveals a large deformation parameter $\beta_2$ = 0.40(3),  consistent with no-core shell model predictions of a large neutron
  deformation. This deformed double-closed shell at the neutron drip-line opens a new paradigm. 

\end{abstract}


\end{frontmatter}
Helium, the second most abundant element in the universe, has a closed
shell ($Z$ = 2) of protons. The $N$ = 2 closed shell of neutrons makes
$^4$He doubly-magic. However, the conventionally expected doubly-magic
heavier isotope, $^{10}$He, is unbound. The He chain terminates at the
most neutron-rich nucleus, $^8$He, with $N/Z=3$. It has an interesting
structure with four neutrons forming a neutron-skin around a $^4$He core
\cite{Zhukov1994}.  Despite being at the neutron drip-line of the He
isotopes it has a larger two-neutron separation energy than $^6$He
\cite{Brodeur2012}. This stronger binding suggests a possible closed
sub-shell at $N$ = 6 which would make $^8$He a doubly closed shell
nucleus. Our knowledge thus far has shown the handful of doubly
closed-shell nuclei to be spherical. Here we investigate if that holds
true for $^8$He from its inelastic excitation that characterizes
nuclear deformation.

The measured charge radius of $^8$He is smaller than that of $^6$He
\cite{Mueller2007}. This decrease in charge radius compared to the preceding isotope is consistent with other $N$ = 6
isotones \cite{Angeli2013} providing a tantalizing hint of
a sub-shell gap in He, Li and Be. This sub-shell feature has also been
discussed in Li isotopes in terms of spectroscopic studies and neutron
separation energies \cite{Kanungo2008}. The matter radius of $^8$He is
slightly larger than that of $^6$He while both are more extended than
$^4$He \cite{Tanihata_plb85,Tanihata92,Alkhazov2002}. $^6$He exhibits
a two-neutron halo, while four neutrons form the neutron skin in
$^8$He. Reactions of $^6$He and $^8$He on a Au target
\cite{lemasson2011} also demonstrate differences in transfer of
neutron pairs in the two nuclei, indicative of different
configurations.  In order to understand the nature of the potential
sub-shell gap at $N$ = 6 however, a precise knowledge of the low-lying
excited-state(s) in $^8$He is required - which thus far remains
elusive.

The nucleus $^8$He has no bound excited states. Several experiments,
with limited statistics, report unbound states. An initial study using
inelastic proton scattering at 72$A$~MeV identified the first excited
state to be $2^+$ at an excitation energy of 3.57(12) MeV with a width
of $\Gamma=0.50(35)$ MeV \cite{Korsheninnikov1993}. The excitation
energy resolution however was $\sim$ 1.8 MeV FWHM, which renders this
determination of the resonance width ambiguous. In Ref.\cite{Chulkov1995} this data is explained using phenomenological density distributions 
with a quadrupole deformation parameter of 0.3 in an eikonal model analysis. A coupled channel analysis
of the angular distribution with microscopic potentials based on model transition
densities show the data agrees with phenomenological densities predicting
neutron quadrupole transition matrix elements ($M_n$) ranging from 3.65 - 5.0 fm$^2$ \cite{Lapoux2015}. It is 
discussed that the (p,p') scattering at lower energy
will have stronger sensitivity to $M_n$. In contrast,
measurements performed at higher energies (227$A$~MeV) using both
Coulomb excitation \cite{Meister2002} and fragmentation
\cite{Markenroth2001} reported a very narrow state,
considered to be possibly $2^+$,  below 3 MeV, lower than that observed in the
(p,p$^\prime$) experiment. This state overlaps with a very broad
second excited state which was conjectured to be a 1$^-$ excitation.
The result from a $^{10}\mathrm{Be}(^{12}\mathrm{C},^{14}\mathrm{O})$
multi-nucleon transfer reaction however finds the energy of the first excited
state to be in agreement with that from inelastic proton scattering.
In addition, three higher energy resonances at energies of 4.54(25)
MeV, 6.03(10) MeV, and 7.16(4) MeV were reported
\cite{Stolla1996,Bohlen2002}. A similar energy and width of the first
excited state was also reported in studies of the
$t(^6\mathrm{He},p)^8\mathrm{He}$ reaction
\cite{Golovkov2009,Fomichev2009}. However, these works proposed that a
significant contribution from a $1^-$-state close to the two-neutron
threshold, $S_{2n}$ = 2.13 ~MeV \cite{AME2016}, better
describes the data.  A recent measurement of the breakup of $^8$He at
82$A$~MeV \cite{Xiao2012} interpreted the resonance spectrum with
conclusions more in line with Ref.~\cite{Markenroth2001,Meister2002}.

To derive the quadrupole deformation parameter and to resolve the inconsistencies regarding
the dipole resonance and first 2$^+$ state, this Letter reports the
first low-energy ($\sim$8.25$A$ MeV) measurement of proton inelastic
scattering with high statistics and high energy resolution.
The experiment was performed at the charged particle spectroscopy
station IRIS at TRIUMF in Canada \cite{Kanungo2014}. The $^8$He
nuclei, produced from the spallation of a SiC target with a 500 MeV proton
beam, were re-accelerated to an energy of 8.25$A$ MeV by
the superconducting linear accelerator \cite{Laxdal} and transported
to the ISAC-II experimental hall where the IRIS facility is
located. The beam had an average intensity of $\sim 10^4$~pps and a
purity of $80-90~\%$ at IRIS. The beam impurity was $^8$Li, identified
event-by-event from energy-loss measured using a low-pressure
ionization chamber, operated with isobutane gas at 19.5 Torr, at the
entrance of the experiment setup. Following this, the beam impinged on
a $~100~\mu$m solid H$_2$ target formed on a $4.5~\mu$m Ag backing
foil cooled to $\sim 4~$K. The target cell was surrounded by a copper
heat shield cooled to $\sim 28~$K.

\begin{figure}
\includegraphics[width=0.45\textwidth]{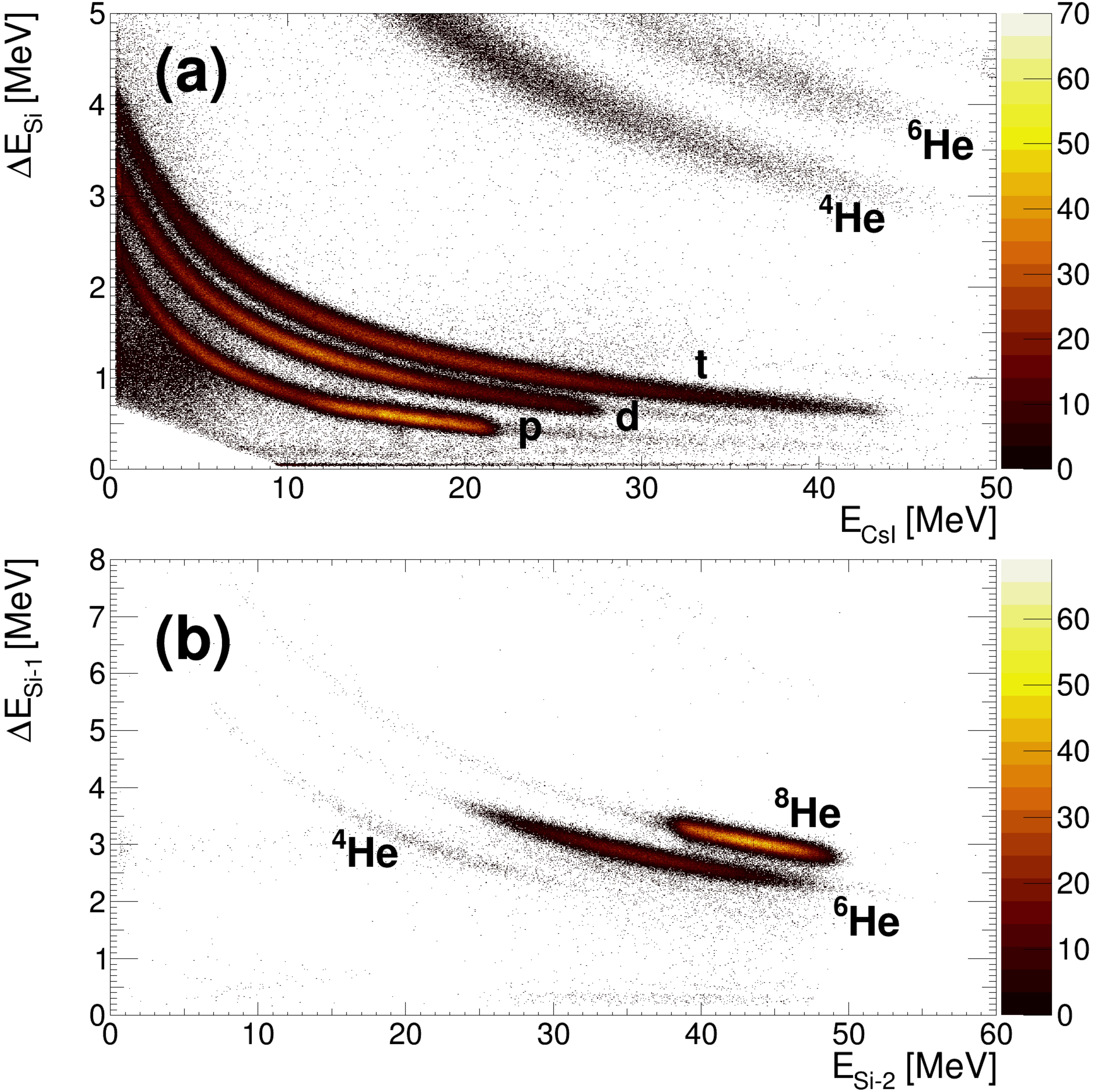}
\caption{\label{fig:1} $\Delta E-E$ identification plots from
  (a) Telescope 1 detecting $p,d,t$ and $^{6,8}$He and (b) Telescope 2 
  detecting $^{6,8}$He ions.}
\label{fig:pid}
\end{figure}

The target-like reaction products, protons ($p$), deuterons ($d$) and
tritons ($t$), as well as helium nuclei were detected and identified
using an array of 100$~\mu$m thick segmented silicon strip detectors
and a 12~mm thick CsI(Tl) array behind it. This $\Delta E-E$ telescope
(Telescope 1) was placed 12.5~cm downstream of the target, covering
laboratory angles of 21-46$^{\circ}$. The top panel of
Fig. \ref{fig:pid} shows the identification plot using this telescope
showing the $p$, $d$ and $t$ loci clearly separated. A second $\Delta
E-E$ telescope (Telescope 2), consisting of  $60~\mu$m and 1~mm
annular double-sided silicon strip detectors, was used to detect the
beam-like He and Li nuclei. Telescope 2, placed 18~cm from the
target, covered scattering angles of $3-10^{\circ}$. The
identification plot of the beam-like He nuclei in coincidence with 
proton detection by Telescope 1 is shown in the bottom panel of Fig. \ref{fig:pid}. The
$^{6,8}$He events are clearly distinguished.

The H$_2$ target thickness was measured from the energy
difference without and with the H$_2$ target using the downstream
telescope. This was done from the peaks of the energy distributions of
both $^8$He and $^8$Li nuclei scattered off the silver foil. In
addition, a silicon surface barrier detector was intermittently
inserted into the beam at 0$^\circ$ located at the extreme downstream
end of the setup as another measurement of the target
thickness. Measurements with a warm target cell without hydrogen were
used to estimate the background from fusion-evaporation reactions
coming from the silver foil. The detection efficiency and acceptance
of the telescopes were determined from simulations of the experiment
in which the energies and momenta of the particles were generated
according to phase space decays and which included the experimental
resolution of the detectors.

\begin{figure}[tb]
\includegraphics[width=0.45\textwidth]{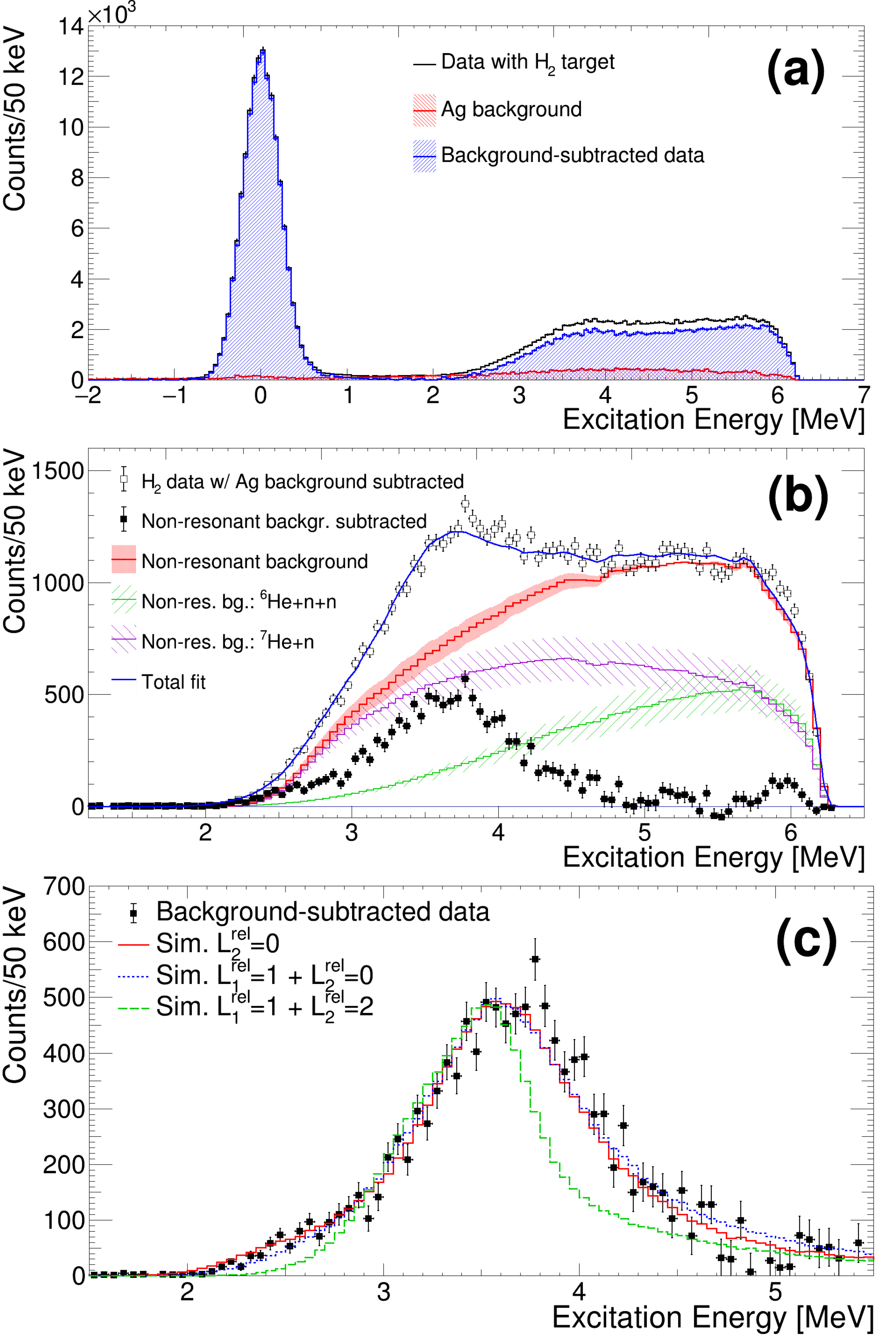}% Here is how to import EPS art
\caption{\label{fig:2} (a) Measured excitation energy spectrum of
  $^8$He. The red / blue histogram shows the measured background
  spectrum from the Ag foil / the spectrum with H$_2$ target after Ag
  foil background subtraction. (b) The background subtracted
  non-elastic excitation spectrum with coincident detection of protons
  and $^6$He. The red shaded band shows the non-resonant background
  from reactions (A) and (B). The individual non-resonant components are shown by the green and magenta curves for channels (A) and (B), respectively.  (c) The observed resonance after
  subtraction of the non-resonant background. The simulated resonance spectra (see text ) for different $L^{rel}$ decay possibilities are shown by the red / blue dotted/green dashed curves labeled in the legend. }
\label{fig:resonance}
\end{figure}

The excitation energy of $^8$He was reconstructed from the measured
energies and angles of the detected protons using the missing mass
technique. The excitation spectrum is shown in
Fig. \ref{fig:resonance}(a). The ground state can be clearly seen and
has negligible background from reactions in the Ag foil. Above the
neutron threshold the excitation spectrum has a strong contribution
from non-resonant reactions together with resonant
excitations. Looking only at events in which $^6$He was detected in
coincidence with scattered protons, the non-resonant background can be
caused by two reactions, $ (\mathrm{A}) \: p + {}^8\mathrm{He}
\rightarrow p + {}^6\mathrm{He} + n + n $ and $ (\mathrm{B}) \: p +
            {}^8\mathrm{He} \rightarrow p + {}^7\mathrm{He} + n $.

The non-resonant reaction kinematics were simulated considering isotropic emission of the reaction products in the center of mass frame. The simulation includes detector geometrical acceptance and resolution effects. The resulting energy of the protons from these non-resonant channels was used to construct the excitation energy spectrum of $^8$He in the identical process of missing mass technique as adopted for the (p,p') inelastic scattering reaction channel. The measured non-elastic spectrum (Fig.~\ref{fig:resonance}(b)) was fitted with a sum of non-resonant channels (A) and (B) with their amplitudes as free fit parameters and a simulated resonance with Voigt function profile where the resonance energy, the width and amplitude were free parameters in the fit. The resonance width contains decay angular momentum energy dependence. 

The blue curve in Fig.~\ref{fig:resonance}(b) shows
the best obtained fit. The red hatched area denotes the contribution
with uncertainty by the non-resonant background. The overall total
strength of the non-resonant contributions from reactions (A) (green curve, Fig.~\ref{fig:resonance}(b))  and (B) (magenta curve, Fig.~\ref{fig:resonance}(b))
 was determined by the resulting best fit parameters considering non-resonant and resonant contributions to the total spectrum. The hatched band indicates the uncertainty. This leads to the non-resonant phase space describing the high energy end of the spectrum. We have not assumed any high excitation energy resonance since there is no clear resonance peak observed in this region. Theoretical predictions of, the 1$^+$ excited state energy by continuum shell model \cite{Volya2006}  and the no core shell model discussed below is $\sim$ 6 MeV which is at the limit of our detection. The differential cross sections for the total spectrum and the derived non-resonant backgrounds can be found in Fig.1(Sup) of the Supplementary Material. 
 
The excitation spectrum after subtraction of the non-resonant
background is shown in Fig. \ref{fig:resonance}(c). In the configuration of the $^8$He(2$^+$) state only the component with core $^6$He$_{gs;0^+}$ decays. The possible decay branches can be one neutron emission to the $^7$He+n threshold and two-neutron emission to the $^6$He$_{gs}$+$nn$ threshold. For $^8$He(2$^+$) = $^6$He$_{gs}$(0$^+$)+$nn$, since the combined intrinsic spin of neutron-neutron ($nn$) cluster $S_{nn}$ = 0, the $nn$ orbital angular momentum $L_{nn}$ = 0, 2 leads to possible  $^6$He-$nn$ relative angular momentum $L^{rel}_2$ = 2, 0 for two-neutron decay to $^6$He$_{gs}$. The relative angular momentum for one-neutron decay to $^7$He+n is $L^{rel}_1$ =1. The spectrum extends below the $^7$He+n threshold signifying decay to $^6$He$_{gs}$+$nn$ to be present. We analyzed the spectrum with a Voigt function with an energy
dependent width ($\Gamma(E) = \Gamma_0\sqrt{(E/E_r)}$) \cite{AlKalanee2013} of the Breit-Wigner resonance profile. $E_r$ is the resonance energy. This energy dependence corresponds to $L^{rel}_2$ = 0 (Fig.~\ref{fig:resonance}(c) red curve). The resultant reduced chisquare from the fit is 1.43. We also performed a fit of the data considering a single resonance state to decay by sum of  $L^{rel}_1$ = 1 and $L^{rel}_2$ = 0, resulting in reduced chisquare value of 1.83 (Fig.~\ref{fig:resonance}(c) blue dotted curve). The similarity of the two fits suggests that the effects of detector acceptance and resolution probably masks a clear distinction. The sum of  $L^{rel}_1$ = 1 and $L^{rel}_2$ = 2 (Fig.~\ref{fig:resonance}(c) green dashed curve) does not explain the data having a reduced chisquare value of 6.9. The narrower width for the $L^{rel}_2$ =2 curve is due to its smaller penetrability.

The determined position and intrinsic width of the resonance from the red curve (Fig. \ref{fig:resonance}(c)) is $E^* = 3.53(4)$~MeV
and $\Gamma = 0.89(11)$~MeV, respectively. The resonance peak from the blue dotted curve (Fig. \ref{fig:resonance}(c)) is 3.56(4) MeV which is in agreement with that from the red curve.  The average resonance energy derived from the two fits is 3.54(6) MeV. The excitation energy
resolution was 0.15~MeV ($\sigma$) at an excitation energy of 3.5~MeV
as determined from simulations which were consistent with the elastic
scattering peak width. The excitation energy is in agreement with the
previous measurements using inelastic scattering and transfer
reactions. However, this high resolution measurement defines precisely
the resonance width which agrees only with the upper uncertainty end
of that reported in Ref.~\cite{Korsheninnikov1993}.  Including an
additional resonance in the fit does not improve the description of
the data. 
The angular distribution is not consistent with a dipole excitation and hence does not align with the conclusion from the breakup
experiments \cite{Markenroth2001,Meister2002,Xiao2012}. The (p,p') and (d,d') reactions can populate low-lying dipole resonance states as seen in Refs.\cite{Kanungo2015,Tanaka2017}.
This suggests that the breakup reactions likely exhibit strong non-resonant dipole
transitions to the continuum, as in $^{11}$Be 
\cite{Fukuda2004} and $^8$B \cite{Davids2001} breakup. 

We mention here that a fit to the full spectrum with the sum of non-resonant background channels and two separate resonance states described by Voigt functions with $L^{rel}_1$ = 1 and $L^{rel}_2$ = 0 whose peak positions and widths are free parameters in the fit results in a reduced chisquare value of 2.76. The spectrum fit and the angular distributions resulting from the two different resonance peaks are included in the Supplementary Material (Fig.2(Sup)) to show that neither of them are consistent with a dipole excitation. 

\begin{figure}[tb]
\includegraphics[width=0.45\textwidth]{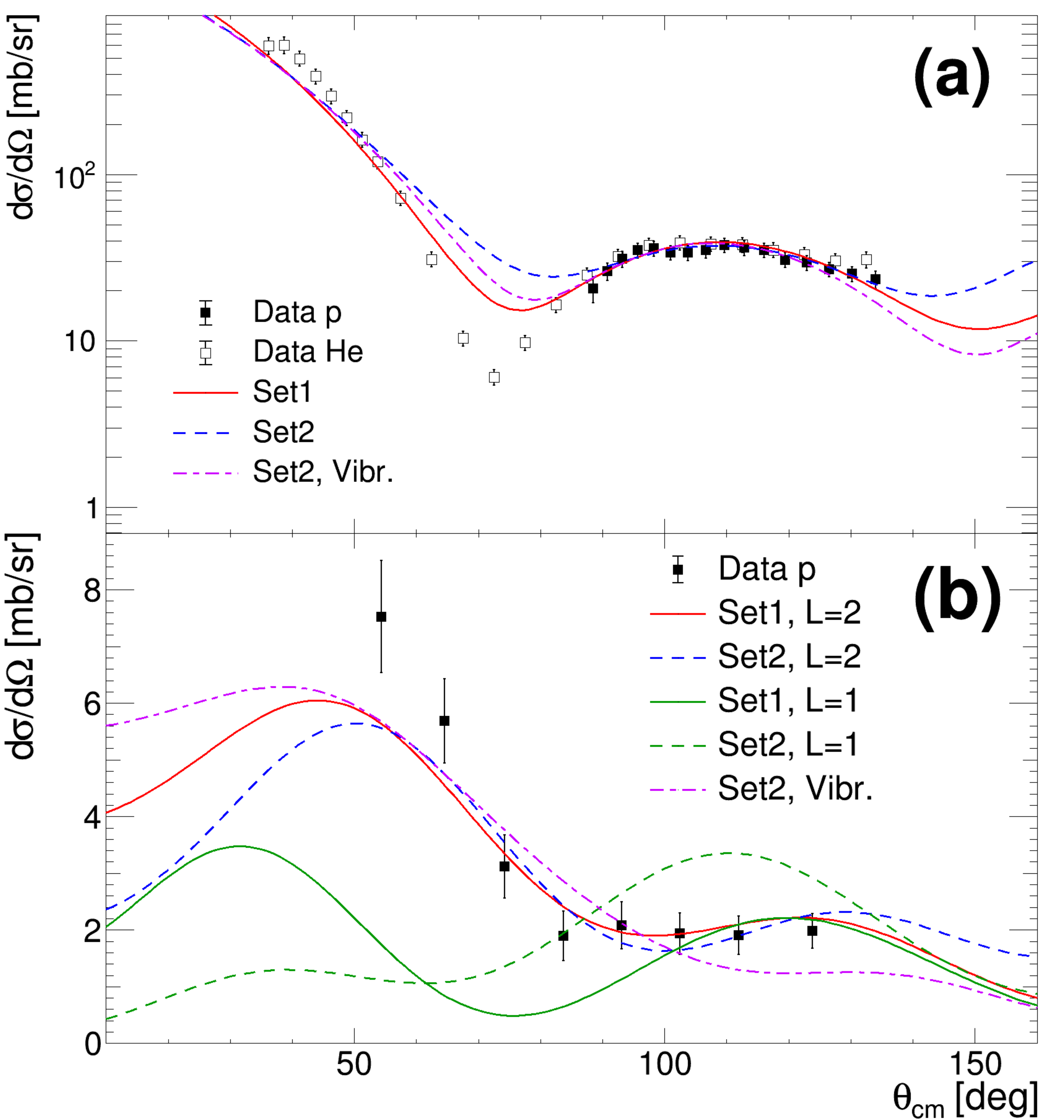}% Here is how to import EPS art
\caption{\label{fig:3} Differential cross sections in the
  center-of-mass frame for elastic (a) open/closed symbols are from detection of $^8$He/$p$ and resonant inelastic scattering
  (b). The curves show CC and DWBA calculations. The red solid  /
  blue dashed  curves are with optical potential
  Set 1/Set 2 and with $L$=2 excitation and rotational model in (b). The pink dashed-dotted curve shows CC calculation for $L$=2 excitation with Set 2 and vibrational model. The green solid/dashed curve in (b) is for $L$=1 excitation with potential Set1/Set2.}
\label{fig:diffCS}
\end{figure}

Differential cross sections in the center-of-mass system for elastic
scattering are shown in Fig.~\ref{fig:diffCS}(a) and for the excited
state at 3.54 MeV in Fig.~\ref{fig:diffCS}(b).  The cross sections were
obtained from the background subtracted spectra where the non-resonant
contribution for inelastic scattering is subtracted as well. The excited state cross section is obtained over the entire excitation range with counts in the resonance profile extracted from the data as described above. The elastic scattering cross section was also obtained from detection of the
scattered $^8$He only (Fig.~\ref{fig:diffCS}(a) open symbols). In order
to determine the $p + ^8$He optical potential parameters, these two
angular distributions were fitted simultaneously with coupled-channel (CC) and
one-step distorted wave Born approximation calculations (DWBA) using the code
SFRESCO \cite{Thompson1988} to obtain the best-fit solution. The DWBA calculations with collective form factor use the rotor model which in first order is same for vibrational model. The CC calculations with rotor model and vibrational model give slightly different fits ((Fig.~\ref{fig:diffCS}). The deformation length $\delta$ was included in the fit to describe
the inelastic scattering data containing the imprint of
deformation. The inelastic scattering angular distribution is
explained by a quadrupole transition, $L = 2$, to the first excited
state.  

We derived two sets of optical potential parameters (Table 1) that
describe the data with DWBA (Set 1) and CC (Set 2) calculations for the entire angular range of elastic scattering. 
The fits to the data using them are shown in Fig.~\ref{fig:diffCS}.
It is noteworthy that the sets require large deformation lengths
$\delta_2^{ex}$ = 1.24 - 1.40~fm within the adopted reaction model to explain the inelastic
scattering data within 2$\sigma$ lower uncertainty for $\theta_{cm} <$ 70$^{\circ}$.  The $\delta_2^{ex}$ derived in this framework is found to be consistent with microscopic reaction model calculations presented below. 
Considering the measured matter radius of
$^8$He \cite{Ozawa2001} they correspond to a large quadrupole
deformation parameter of $\beta_2$ = 0.40(3) showing that $^8$He has a deformed sub-shell gap at $N$ = 6.  

The collective vibrational model form factor cannot distinguish between static and dynamic deformation. However, the no-core shell model (NCSM) calculations reported below shows large neutron deformation in the 2$^+$ state of $^8$He. The large neutron quadrupole moment for the 2$^+$ state predicted by the NCSM calculations suggest $^8$He as a nucleus with a significant intrinsic deformation in contrast to a spherical (vibrational) picture, for which the 2$^+$ reorientation term would vanish. The microscopic transition density obtained in the no-core shell model leads to a quadrupole deformation length consistent with that derived from the collective form factor as discussed below. 

\begin{table*}[]
    \centering
    \caption{Optical potential parameters for $^8\mathrm{He}+p$, determined from a simultaneous fit to the elastic and inelastic scattering data. The depth, radius, diffuseness parameters for the real potential are $V, r, a$ and for the surface imaginary potential are $V_s, r_s, a_s$, respectively. The degrees of freedom ($dof$) were 30.}
    \begin{tabular}{cccccccccc}
       \hline
                && $V$~[MeV] & $r$~[fm]  & $a$~[fm]  & $W_s$~[MeV]   & $r_s$~[fm]    & $a_s$~[fm]    & $\delta$~[fm] & $\chi^2/dof$\\
         \hline
        Set 1   &DWBA& 46.3      & 1.65      & 0.35      & 22.8          & 1.77          & 0.27          & 1.40          & 1.60\\
        Set 2   &CC - Rotor Model & 50.5      & 1.51      & 0.33      & 20.2          & 1.79          & 0.19          & 1.24          & 1.36\\
        Set 2   &CC - Vibrational Model & 50.5      & 1.51      & 0.33      & 20.2          & 1.79          & 0.19          & 1.32          & 1.60\\
         \hline
    \end{tabular}
    \label{tab:opm}
\end{table*} 

The $\beta_2$ values for heavier $N$ = 6 isotones, are 1.14(6) for
$^{10}$Be and 0.582(24) for $^{12}$C \cite{RA01} which suggests that
deformation persists from stable nuclei to the neutron-rich
region. The excitation energies of the 2$^+$ states for the $N$ = 6
isotones are similar, 4.44 MeV in $^{12}$C, 3.37 MeV in $^{10}$Be and
3.54 MeV in $^{8}$He.  Therefore, the extent of the sub-shell gap at
$N$ = 6 may be similar along the isotonic chain but it becomes
prominent towards the neutron-drip line due to the disappearance of
the strong shell closure at $N$ = 8. In $^{14}$O however, the 2$^+$
state lies at a much higher excitation energy of 6.6 MeV, reflecting
the presence of protons in the filled 1$p_{1/2}$ orbital, causing a
wider gap at $N$ = 6 due to the attractive proton-neutron tensor
force. The question remains open regarding the extent of deformation
in heavier nuclei with closed neutron sub-shells, such as neutron-rich
Ca isotopes, where some theoretical predictions suggest the
drip-line extending to $^{72,74}$Ca \cite{PhysRevLett.122.062502,meng2002}.

A small part of the resonance spectrum extends below the $^7$He+n threshold indicating that decay to the $^6$He$_{gs}$+2n threshold is important. The measured angular distribution is not supportive of a low-energy dipole resonance ($L$ = 1, Fig.~\ref{fig:diffCS}(b)). Further studies may aid in a complete understanding of this feature. 

\begin{figure}[tb]
\includegraphics[width=0.45\textwidth]{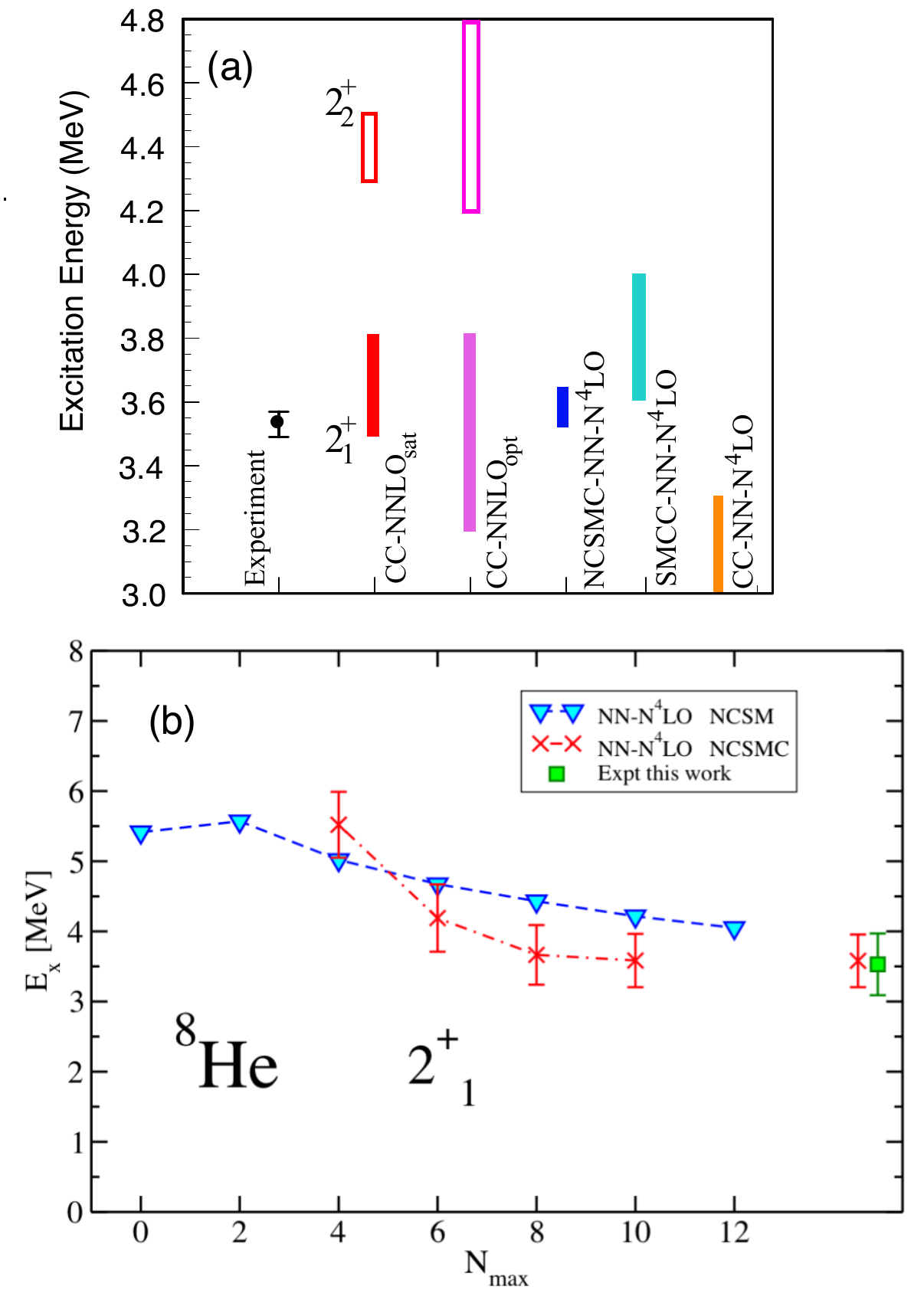}
\caption{\label{fig:4} (a) Comparison of the observed 2$^+$ excited state
  in $^8$He to \textit{ab initio} predictions. The red bands show
  results using the NNLO$_{\rm sat}$ interaction with the EOM-CCSDT-3
  method for the 2$_1^+$ and 2$_2^+$ states. The pink bands
  show results for the 2$_1^+$ and 2$_2^+$ states with the NNLO$_{\rm
    opt}$ interaction using the EOM-CCSDT-3 and SMCC methods. The blue
  band shows the result for the 2$_1^+$ state with the NN-N$^4$LO
  interaction in a NCSMC calculation, no 2$_2^+$ state is found within
  this excitation energy range. The cyan / orange bands show
  results with the SMCC / EOM-CCSDT-3 methods using the NN-N$^4$LO
  interaction. (b) The $2^+_1$ excitation energy dependence of $^8$He on the basis size for the NCSM and NCSMC calculations with the NN-N$^4$LO interaction. Extrapolated values and the data are shown on the right. The vertical bars represent resonance widths obtained in the NCSMC calculations and in the experiment.}
\label{fig:abinitio_comp}
\end{figure}

For theoretical descriptions of the 2$^+$ states in $^8$He, we
employed two many-body approaches, namely coupled-cluster theory and
the no-core shell model (with continuum) using several chiral
interactions. The measured energy of the resonance (2$_1^+$
state) is compared to the {\it ab initio} calculations in
Fig.~\ref{fig:abinitio_comp}.

We used two different coupled-cluster methods~\cite{Hagen2014}.
First, for the chiral interaction NNLO$_{\rm sat}$~\cite{ekstrom2015}
we employ the equation-of-motion technique with up to
three-particle--three-hole (3p-3h) excitations (so called
EOM-CCSDT-3)~\cite{watts1996} of the closed-shell $^8$He reference
state, and find the 2$_1^+$ state at an excitation energy of 3.5 to
3.8 MeV. The range reflects model-space uncertainties.
The Hartree-Fock basis is built from model spaces consisting of 11 to
15 oscillator shells with frequencies between 12 and 16~MeV. We find
that $^8$He is bound by almost 3~MeV with respect to $^4$He, in
agreement with data ($S_{4n} \sim 3.11$~MeV). Second, we employ the
chiral nucleon-nucleon (NN) interaction NNLO$_{\rm
  opt}$~\cite{ekstrom2013}. For this interaction $^8$He is not bound
and about 1.5~MeV above the $^4$He ground-state energy.  We again used
the EOM-CCSDT-3 approach and also the shell-model coupled-cluster
method (SMCC)~\cite{sun2018}. This method employs a $^4$He core and
constructs a valence-space Hamiltonian in the $0p_{3/2}$, $0p_{1/2}$,
and $1s_{1/2}$ shells based on computations of the $A=5,6$ body
problems in 5 to 13 oscillator shells (and frequencies of 12 to
22~MeV). In the valence space four-neutron correlations are treated
exactly.  The EOM-CCSDT-3 and SMCC methods yield an excited $2_1^+$
state at about 3.2 and 3.8~MeV, respectively, and we take this range
as a systematic uncertainty.  Both predictions agree with the data
within the theoretical uncertainty band shown in
Fig.~\ref{fig:abinitio_comp}(a) with and without the three-nucleon force.

We also applied the no-core shell model (NCSM)~\cite{Barrett2013} to
calculate properties of $^8$He. In the NCSM, the many-body wave
function is expanded over a basis of antisymmetric $A$-nucleon
harmonic oscillator (HO) states. The basis contains up to $N_{\rm
  max}$ HO excitations above the lowest possible Pauli configuration
and depends on an additional parameter $\Omega$, the frequency of the
HO well. We employed the same Hamiltonian as in our recent
investigation of $^9$He~\cite{PhysRevC.97.034314}. The NN interaction,
denoted here as NN-N$^4$LO, is from the fifth order chiral expansion
(N$^4$LO) of Ref.~\cite{PhysRevC.96.024004} and was renormalized by
the SRG approach~\cite{PhysRevC.75.061001} with an evolution parameter
$\lambda_{\rm SRG}{=}2.4$~fm$^{-1}$. The three- and higher-body SRG
induced terms were not included. We performed calculations up to
$N_{\rm max}{=}12$ with $\hbar\Omega{=}20$ MeV. We find $^8$He bound
by about 2 MeV with respect to $^4$He. The NCSM calculations yield a
large quadrupole neutron moment $Q_n = 6.15~e$fm$^2$ and a small
proton quadrupole moment, $Q_p= 0.60~e$fm$^2$ for the 2$^+_1$
state. For $^{12}$C we predict $Q_n{\approx}Q_p{\sim}
6~e$fm$^2$. Thus, the neutron deformation in $^8$He is similar to that
in $^{12}$C and qualitatively consistent with the experimental
observations discussed above. In the Variational Monte Carlo framework $Q_p$ of $^8$C(2$^+$) is 
5.6 $e$fm$^2$ which can reflect the $Q_n$ of $^8$He(2$^+$) considering charge symmetry \cite{Wiringa}.
 
As the $2^+_1$ state is unbound, its excitation energy convergence is
slow in NCSM as seen in Fig.~\ref{fig:abinitio_comp}(b). To improve the theoretical description, we applied the
no-core shell model with continuum
(NCSMC)~\cite{PhysRevLett.110.022505,PhysRevC.87.034326,Navratil2016}. Optimally,
three-body cluster NCSMC~\cite{PhysRevC.97.034332} with $^6$He+n+n or
even five-body $^4$He+4n continuum should be used. That is, however,
beyond our current technical capabilities. As the $^7$He ground state
resonance is rather narrow (150 keV), it is reasonable to use as the
simplest alternative the $^7$He(gs)+n cluster to extend the $^8$He
NCSM basis. This results in a greatly improved convergence of the
$2^+_1$ excitation energy, with the extrapolated value of 3.58(6)
MeV (Fig.~\ref{fig:abinitio_comp}(b)). In addition, we calculate the $2^+_1$ width to be 750(50)
keV. Overall, with this interaction we obtain an excellent agreement
with the present experimental measurement. We note that the only other resonance 
we find in the calculation below 6 MeV in $^8$He excitation energy is a broad $1^+$ state. 
In particular, we do not see any evidence for a $1^-$ resonance in this energy range. 
Both the NCSM and NCSMC calculations were performed using the HO frequency of $\hbar\Omega~{=}~20$ MeV determined as optimal for the $^8$He ground state with the NN-N$^4$LO interaction as illustrated in Fig. 2 in Ref.~\cite{PhysRevC.97.034314}. Compared to that paper, the present NCSM calculations were extended to $N_{\rm max}~{=}~12$ using the importance truncation~\cite{Roth2007,Roth2009}. The calculated $N_{\rm max}~{=}~12$ ground-state energy, -28.2 MeV, is in line with the extrapolation shown in Fig. 4 of Ref.~\cite{PhysRevC.97.034314}.
 
We also benchmarked the coupled-cluster computations with the NCSM
using the NN-N$^4$LO potential.  For the ground-state energy, the
extrapolated NCSM result \cite{PhysRevC.97.034314} is
$E=-30.23(30)$~MeV, while SMCC and CCSDT-3 yield $-30.3$ and
$-29.0(5)$~MeV, respectively. For the 2$^+_1$ state, we find
excitation energies of $3.8(2)$ and $3.1(2)$~MeV for SMCC and
EOM-CCSDT-3, respectively. The SMCC results agree with the NCSM, while
CCSDT-3 and EOM-CCSDT-3 are less accurate.  We note that the
EOM-CCSDT-3 calculations yield two nearby $2^+$ states at $3.1(2)$ and
$3.8(2)$~MeV. The first state carries only about 40\% of $1p-1h$
amplitudes from the reference state, while the second state exhibits
about 70\% of $1p-1h$ amplitudes.  Thus, the EOM computations are not
converged with respect to wave function correlations, and $^8$He is
not a closed-shell nucleus for the employed potential.

\begin{figure}[tb]
\includegraphics[width=0.45\textwidth]{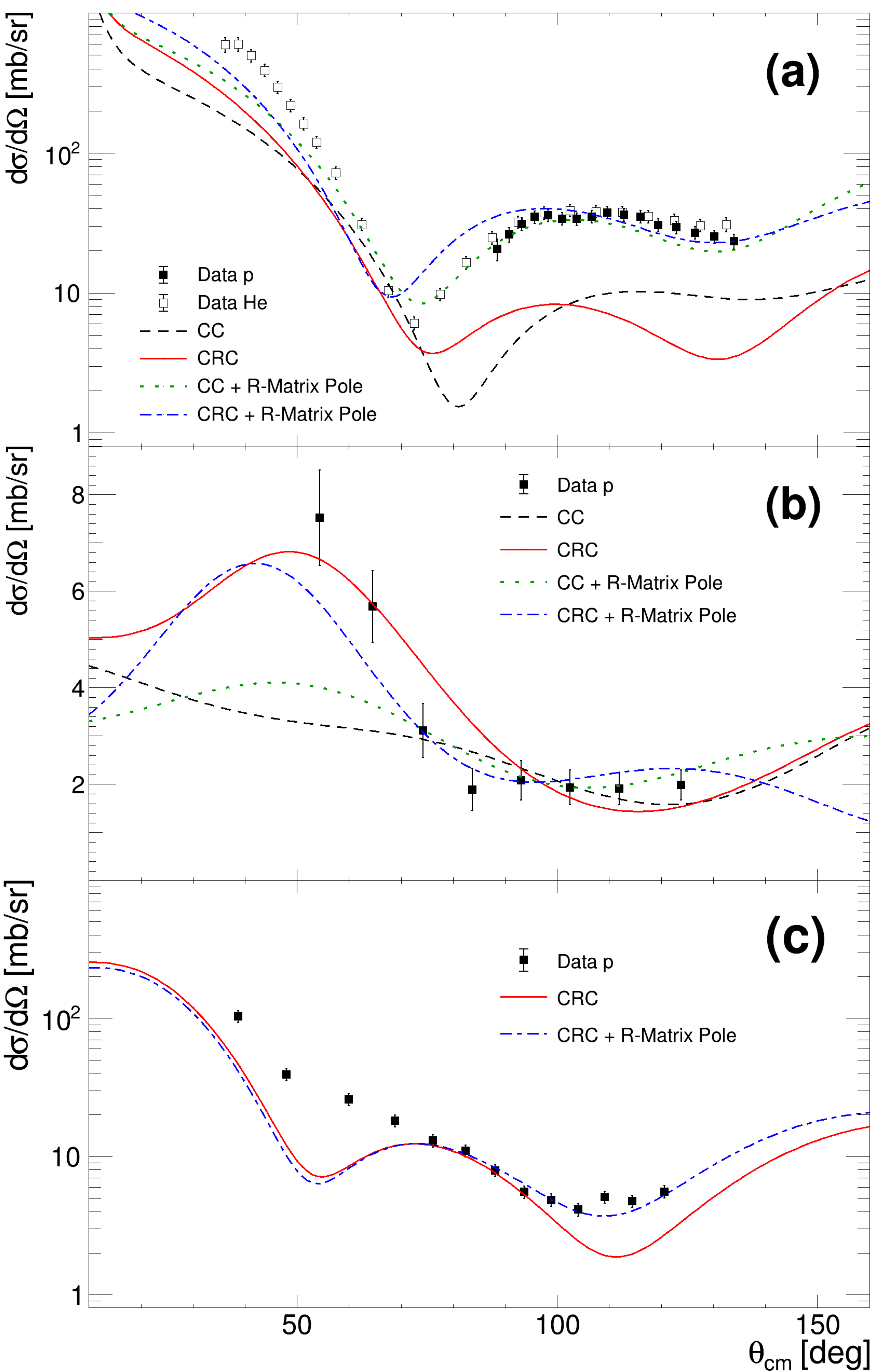}
\caption{\label{fig:5} The measured differential cross sections for (a) $^8$He(p,p) (b) $^8$He(p,p')$^8$He(2$^+$) and (c) $^8$He(p,d)$^7$He shown by the symbols. The black dashed / red solid curves are CC / CRC calculations with NCSM densities (see text). The blue dotted / dashed-dotted curve shows CC / CRC calculation including an R-matrix pole. }
\label{fig:CRCDiffCross}
\end{figure}

The differential cross section are also analyzed in terms of coupled-channels (CC) and coupled-reaction-channels (CRC) calculations using whenever possible structure inputs from the aforementioned NCSM calculations. In the CC calculations, only the the elastic and inelastic channels were considered. The p+$^{8}$He(g.s.) and p+$^{8}$He(2$^+$) diagonal potentials as well as the quadrupole coupling between these two channels were computed by a single folding procedure, convoluting the JLMb interaction of Ref.~\cite{PhysRevC.58.1118} with the NCSM  matter and transition densities.   The result of this calculation, shown by the black dashed curve in Fig.\ref{fig:CRCDiffCross}, does not describe well the shape of the data. In the CRC calculations, in addition to the couplings considered in the CC calculations, we included the coupling to the d+$^{7}$He(g.s.) and p+n+$^{7}$He(g.s.) channels, the latter accounting for the deuteron continuum. The required $\langle {\rm ^7He} | {\rm ^8He(g.s.)}\rangle$ overlap function was approximated by a single-particle wavefunction calculated in a Woods-Saxon potential with the depth adjusted to give the experimental separation energy and the geometry adjusted to reproduce the NCSM overlap in the interior. The same Woods-Saxon geometry was adopted for the   $\langle {\rm ^7He} | {\rm ^8He(2^+)}\rangle$ overlap. The results of these calculations are given by the red solid lines in Fig.\ref{fig:CRCDiffCross}. The agreement with the inelastic and transfer angular distributions is rather satisfactory, but not for the elastic scattering scattering.  The coupling  between the $p$+$^{8}$He(2$^+$) channel and the d + $^{7}$He channel was found to be very important in reproducing the shape of the inelastic scattering data.  It is interesting to note that the deformation parameter derived from the NCSM quadrupole transition density that explains the data is found to be $\delta_2$ = 1.39~fm, in good agreement with that derived from the fit to the experimental inelastic cross sections using the collective model with a deformed Woods-Saxon potential discussed above. 

The incomplete description of the elastic scattering data (Fig.\ref{fig:CRCDiffCross}) with CC and CRC could be due to the effect of compound nucleus resonance(s) in $^{9}$Li. Although a detailed investigation of this effect is beyond the scope of the current work, to highlight the possible effect we have performed  CC and CRC calculations including an R-matrix pole representing the effect of a compound-nucleus resonance with a $J^\pi=5/2^+$ and with the energy and reduced width amplitudes adjusted to reproduce in the best possible way the measured elastic and inelastic data. As a result, we obtain a formal energy of $E_\mathrm{c.m.} \simeq 5.5$~MeV ($E_x=19.4$~MeV with respect to the $^{9}$Li(g.s.)). The resultant elastic and inelastic distributions are given by the blue dotted curve for CC + R matrix pole and blue dashed-dotted curve for CRC + R matrix pole in Fig.\ref{fig:CRCDiffCross}. It is seen that the inclusion of such resonance can result in a significantly improved agreement with the data.

In summary, a measurement of proton inelastic scattering of $^8$He at 8.25$A$ MeV affirms the first excited state to be an unbound 2$^+$
state at an excitation energy of 3.54(6) MeV with a width of 0.89(11) MeV (FWHM).  
Analysis of the measured angular distribution yields a quadrupole deformation parameter of $\beta_2$ = 0.40(3). The deformation length is consistent with calculations in a no-core shell model framework. Microscopic CRC calculations with NCSM densities explain the inelastic scattering yielding a deformation length 1.39 fm, providing further support for the large deformation. {\it Ab initio} calculations in a coupled cluster framework and NCSMC find a 2$^+_1$ excitation energy in good agreement with the data. The resonance width predicted by the NCSMC is also consistent with the data. The high-quality data, signalling deformation at the $N$ = 6 drip-line, open exciting prospects for further investigations.

The authors thank the TRIUMF ISAC beam delivery team and the cyclotron
operations team. The support from NSERC, Canada Foundation for Innovation and Nova Scotia Research and Innovation Trust is gratefully
acknowledged. TRIUMF receives funding via a contribution through the
National Research Council Canada. The support from RCNP for the target
is gratefully acknowledged. It was partly supported by the
grant-in-aid program of the Japanese government under the contract
number 23224008 and 14J03935. The use of the S3 detector provided by
C.Y. Wu of LLNL is much appreciated. Computing support from Westgrid
and Compute Canada is gratefully acknowledged. This work was supported
by the Office of Nuclear Physics, US Department of Energy, under
grants DE-FG02-96ER40963 and DE-SC0018223, and at Lawrence Livermore
National Laboratory under Contract DE-AC52-07NA27344, as well as by
the Field Work Proposal ERKBP72 at Oak Ridge National Laboratory
(ORNL). Computer time was provided by the Innovative and Novel
Computational Impact on Theory and Experiment (INCITE) programme. This
research used resources of the Oak Ridge Leadership Computing Facility
located at Oak Ridge National Laboratory, which is supported by the
Office of Science of the Department of Energy under contract
no. DE-AC05-00OR22725.  J.A.L. and A.M.M. are  supported by the Spanish Ministerio de Ciencia, Innovaci\'on y Universidades (including FEDER funds)  under project FIS2017-88410-P and by the European Union's Horizon 2020 research and innovation program under Grant Agreement No.\ 654002. Discussions with R.B. Wiringa is gratefully acknowledged. 
\section*{References}
%\bibliography{references_rev2}

\end{document}